%% file: Manuscript.tex
\documentclass[reprint, superscriptaddress, amsmath,amssymb, aps, floatfix]{revtex4-2}

\usepackage{graphicx}
\usepackage{dcolumn}
\usepackage{hyperref} 
\usepackage{bm}
\usepackage{dsfont}
\usepackage{xcolor}
\usepackage{hhline} 

\newcommand{\xUZH}{Department of Physics, University of Z{\"u}rich, 8057 Z{\"u}rich, Switzerland}

\begin{document}

\preprint{APS/123-QED}

\title{Simulating fermionic fractional Chern insulators \\
with infinite projected entangled-pair states}

\author{Hao Chen}
\affiliation{\xUZH}

\author{Titus Neupert}
\affiliation{\xUZH}

\author{Juraj Hasik}
\affiliation{\xUZH}

\date{\today}

\begin{abstract}
Infinite projected entangled-pair states (iPEPS) provide a powerful variational framework for two-dimensional quantum matter and have been widely used to capture bosonic topological order, including chiral spin liquids. Here we extend this approach to \emph{fermionic} topological order by variationally optimizing $U(1)$-symmetric fermionic iPEPS for a fractional Chern insulator (FCI), with bond dimensions up to $D=9$. We find evidence for a critical bond dimension, above which the ansatz faithfully represents the FCI phase. The FCI state is characterized using bulk observables, including the equal-time single-particle Green's function and the pair-correlation function, as well as the momentum-resolved edge entanglement spectrum. To enable entanglement-spectrum calculations for large iPEPS unit cells, we introduce a compression scheme and show that the low-lying part of the spectrum is already well converged at relatively small cutoff dimensions.
\end{abstract}

\maketitle

\emph{Introduction---}Fractional Chern insulators (FCIs) are lattice analogues of the fractional quantum Hall (FQH) states realized without external magnetic fields \cite{Neupert2011,Regnault2011, sheng2011fractional} in fractionally
filled narrow Chern bands~\cite{Sun2011,Tang2011}. 
As a chiral topologically ordered phase, FCIs host fractionalized anyonic excitations and chiral gapless edge modes.
Recent experimental realizations of this phase of matter in two-dimensional (2D) moiré materials \cite{cai2023signatures,park2023observation,zeng2023thermodynamic,lu2024fractional} have reignited the interest in FCIs.
Simulating microscopic Hamiltonians that host FCIs is therefore crucial---both to test theory and to connect with experimental observations. 
So far, the primary numerical methods employed in this field have been restricted to exact diagonalization (ED) \cite{Neupert2011,Regnault2011, sheng2011fractional,bergholtz2013topological} and infinite density matrix renormalization group (iDMRG) \cite{Grushin2015,He2017}, which are limited to small clusters or finite-width cylinders and hence are subjected to finite-size effects.

Infinite projected entangled-pair state (iPEPS), a two-dimensional tensor network ansatz, offers a complementary route: by construction, it is formulated directly in the thermodynamic limit. Despite the no-go theorem forbidding an exact finite-correlation-length iPEPS representation of a gapped chiral phase~\cite{Dubail2015}, iPEPS has proven effective in practice for simulating bosonic chiral topological orders, including chiral spin liquids (CSLs)~\cite{Poilblanc2017chiral, Chen2021, Juraj2022, Sen2022,Budaraju2024,chen2025simulating}, bosonic FQH states \cite{Weerda2024FQHE}, and non-Abelian generalizations~\cite{chen2018non,Sen2024}. Nevertheless, iPEPS has not yet been applied to \emph{fermionic} topological order to the best of  our knowledge. 

Recent algorithmic advances have made iPEPS a powerful toolkit for tackling this problem.
First, we leverage standard $U(1)$-symmetric iPEPS~\cite{McCulloch2007,Singh2010,Bauer2011symmetry,Singh2011} to implement fermionic charge conservation, which enables efficient encoding of fermionic statistics~\cite{Corboz2009,Corboz2010} 
and also yields block-sparse tensors that substantially reduce computational cost, enabling optimizations at larger bond dimensions. 
Second, we use automatic differentiation (AD) \cite{Liao2019} that facilitates straightforward gradient-based optimizations. Unlike the standard imaginary–time evolutions, the gradient-based approach can reliably find long-range entangled phases such as CSLs and the FCI studied here. 
The key new advance comes from the fixed-point AD method~\cite{Liao2019,Boris2022,naumann2024introduction,Anna2025}. 
When combined with a robust gauge-fixing scheme \cite{Anna2025}, we can overcome the performance bottleneck in simulating chiral states: 
the corner transfer matrix renormalization group (CTMRG)~\cite{nishino1996corner,nishino1997corner,Orus2009,Corboz2014,Fishman2018},  
often adopted to compute observables for iPEPS, regularly requires a large number of iterations ($O(10^3)$ for the FCI state studied here) to converge due to the long correlation lengths implied by the no-go theorem. This leads to large memory requirements when backpropagating by AD.
Fixed-point AD circumvents this by directly differentiating the converged fixed point of CTMRG and eliminates the need for storing any intermediates throughout CTMRG.

In this Letter, we extend the applicability of iPEPS to \emph{fermionic} FCIs by integrating the above advances. We carry out gradient optimization on a $U(1)$-symmetric fermionic iPEPS for a spinless-fermion Hamiltonian on the honeycomb lattice and are able to the push the bond dimension of iPEPS to $D=9$. 
We find that a faithful representation of the FCI requires $D\ge 7$, and identify the phase via the equal-time single-particle Green's function, the pair-correlation function, and the edge entanglement spectrum (ES). To access the ES for large iPEPS unit cells, we further introduce a simple compression scheme, which makes the calculations computationally feasible.

\begin{figure}[t]
    \centering
    \includegraphics[width=\columnwidth]{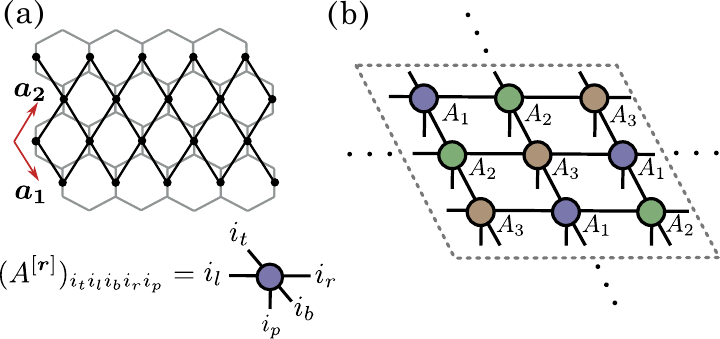}
    \caption{(a) Distinct hopping processes of the Haldane model given by  Eq.~\eqref{eq:Hamiltonian} are shown in different colors, with the arrows for 2nd-NN hopping indicating the phase convention. It is reduced to the corresponding Bravais lattice  by grouping the two sublattice sites A and B within each honeycomb unit cell into a single effective site.
    (b) iPEPS is constructed on the Bravais lattice from rank-5 tensors $A^{[\boldsymbol{r}]}$ defined in Eq.~\eqref{eq:onsiteA}, where the arrows indicate the directions of the $U(1)$ charge flow.
    We parametrize the ground state with three inequivalent tensors $(A_1,A_2,A_3)$ arranged in $3\times3$ unit cell  which periodically tiles the lattice. 
    }
    \label{fig:model}
\end{figure}

\emph{Model and methods---}We study spinless fermions on the honeycomb lattice governed by the Haldane model Hamiltonian~\cite{Haldane1988} with density-density interactions 
\begin{equation}\label{eq:Hamiltonian}
    \hat{H} = -\sum_{i,j} t_{ij} \hat{c}_i^\dagger \hat{c}_j
        + V_1 \sum_{\langle ij \rangle} \hat{n}_i \hat{n}_j,
\end{equation}
where $\langle ij \rangle$ denotes nearest-neighbor (NN) sites, $\hat{c}^\dagger_i$ creates a spinless fermion on site $i$ of the honeycomb lattice and $\hat{n}_i =\hat{c}^\dagger_i \hat{c}_i$ measures the site occupation number. Fermions can hop between NN, second-NN, and third-NN sites with hopping amplitudes $t_1, t_2 e^{\pm i\phi},$ and $t_3$, respectively. The sign convention of the acquired phase $\pm \phi$ is illustrated in Fig.~\ref{fig:model}(a). 
We focus on $V_1/t_1=10.0, t_2/t_1 = 0.7, t_3/t_1 = -0.9$, and $\phi=0.35\pi$, for which the non-interacting part of the Hamiltonian features an almost flat Chern band as the valence band.
When this band is fractionally filled to $\nu=1/3$ filling (one fermion per three unit cells; here $\nu=2$ denotes full lattice filling, i.e., one fermion per honeycomb site) the NN interactions in the Hamiltonian give rise to an FCI ground state, as evidenced by iDMRG simulations~\cite{Grushin2015} on thin cylinders. The FCI phase persists upon including second-NN and third-NN interactions~\cite{HaoChen2025}.

Here, we simulate the ground state of this model directly in the thermodynamic limit, by variationally optimizing an iPEPS. 
The iPEPS encodes the amplitudes $\phi_{\vec{\nu}}$ of the ground state  $|\psi\rangle=\sum_{\vec{\nu}}\phi_{\vec{\nu}} \prod_i(\hat{c}_i^\dagger)^{\nu_i}|0\rangle$ with $\nu_i\in\{0,1\}$ as a contraction of an infinite, translationally invariant two-dimensional tensor network. To construct our iPEPS, we first coarse-grain the honeycomb lattice into its Bravais lattice as shown in Fig.~\ref{fig:model}(a), by assigning a local physical Hilbert space $\mathbb{V}_p = \text{span}(|0\rangle, \hat{c}^\dagger_{\mathrm A} |0\rangle,\hat{c}^\dagger_{\mathrm B} |0\rangle,\hat{c}^\dagger_{\mathrm A}\hat{c}^\dagger_{\mathrm B}|0\rangle)$ of dimension $d=4$ to each unit cell. This is equivalent to a two-flavour fermion with one flavour for each sublattice. Next, we assign a rank-5 tensor $(A^{[\boldsymbol{r}]})_{i_t i_l i_b i_r i_p}$ to each Bravais lattice unit cell $\boldsymbol{r}$ 
\begin{equation}
\label{eq:onsiteA}
    A^{[\boldsymbol{r}]}:  \mathbb{V}_t\otimes \mathbb{V}_l\otimes \mathbb{V}_b^*\otimes \mathbb{V}_r^* \otimes \mathbb{V}{_{p}}\to \mathbb C,
\end{equation}
where $\mathbb{V}_{\alpha =  t, l, b, r}$ are \emph{virtual} Hilbert spaces of \emph{bond dimension} $D$ associated with the corresponding legs of $A^{[\boldsymbol{r}]}$, and $\mathbb{V}^*$ denotes the dual space of $\mathbb{V}$. 
Finally, the tensors $A^{[\boldsymbol{r}]}$ are contracted through their virtual legs $(t, l, b, r)$ according to the lattice connectivity, forming the iPEPS shown in Fig.~\ref{fig:model}(b). This ansatz efficiently approximates amplitudes $\phi_{\vec{\nu}}$, with the precision controlled by $D$.

The Hamiltonian in Eq.~\eqref{eq:Hamiltonian} has a global $U(1)$-symmetry associated with the conservation of fermion number $\hat{N}=\sum_i {\hat{n}_i}$, and hence also its $\mathbb{Z}_2$ subgroup corresponding to fermion parity $\hat{P}=(-1)^{\hat{N}}$. 
We exploit this parity symmetry to efficiently encode fermionic exchange statistics using the swap-gate approach \cite{Corboz2009,Corboz2010}) (see Refs.~\cite{kraus2010fermionic,mortier2025fermionic,Gao2025fermionic} for alternative fermionic encodings). With parity symmetry alone, the filling $\nu$ is typically controlled by adding a chemical potential term $\mu \hat{N}$ to Eq.~\eqref{eq:Hamiltonian} and numerically tuning $\mu$ until the desired filling is achieved (approximately). 
Here, instead, we use $U(1)$-symmetric tensors, which allow us to enforce an integer number of fermions per iPEPS unit cell by attaching auxiliary legs to selected tensors (see Appendix~A). This enables us to realize the desired filling fraction exactly by an appropriate choice of iPEPS unit cell.
Specifically, for filling $\nu=1/3$ we parametrize the ground state as $|\psi_{\nu=1/3}\rangle = |\text{iPEPS}(\vec{A}=[A_1,A_2,A_3])\rangle$ with an enlarged $3\times 3$ unit cell made from three independent $U(1)$-symmetric tensors $A_1,A_2,A_3$, as shown in Fig.~\ref{fig:model}(b), and attach an auxiliary leg to one of the three tensors (here $A_1$) to inject a fermion. This construction fixes the average fermion number within the 
$3\times3$ unit cell to three (three $A_1$), yielding $\nu=3/9=1/3$.
Similarly, to simulate the Chern insulator (CI) phase of Eq.~\eqref{eq:Hamiltonian} with $V_1=0$ at filling $\nu=1$, we use a $1\times1$ unit cell parametrization $|\psi_{\nu=1}\rangle=|\text{iPEPS}(\vec{A}=[A_0])\rangle$, where $A_0$ again carries an auxiliary leg.

\begin{figure}[tb]
    \centering
    \includegraphics[width=\columnwidth]{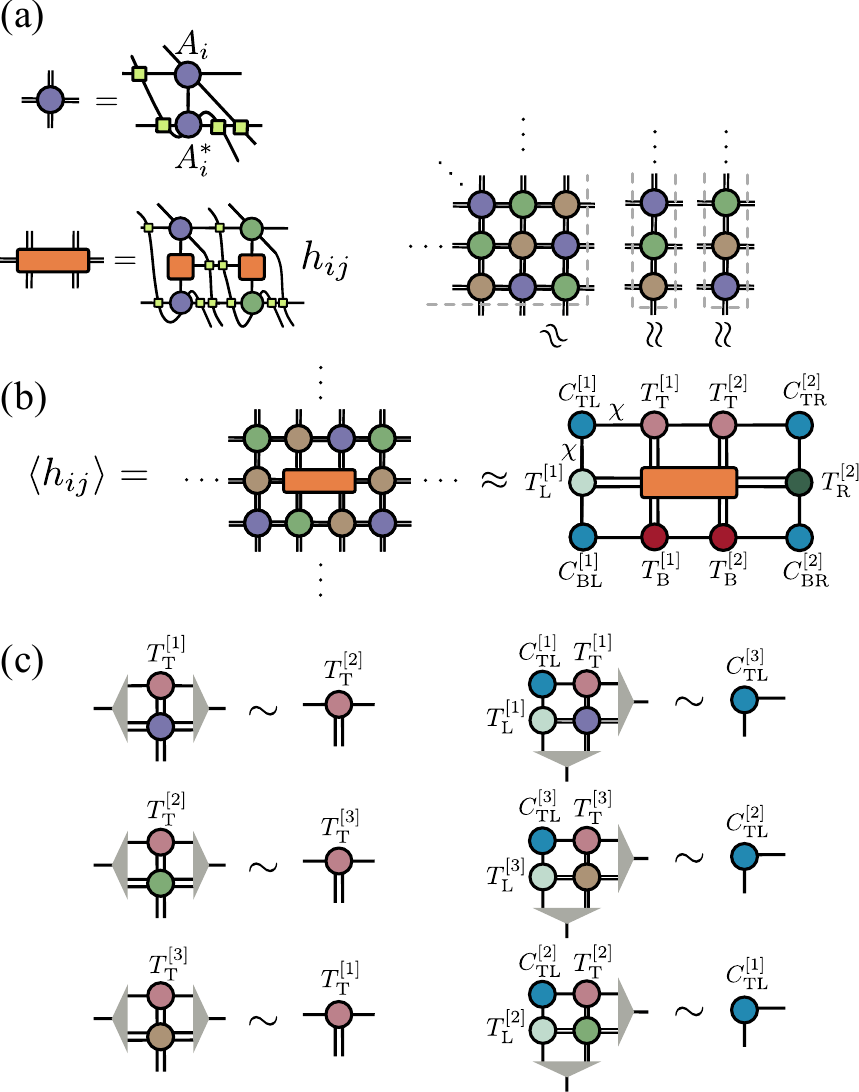}
    \caption{CTMRG routine.
    (a) Double-layer tensors used to evaluate the nearest-neighbor contribution $\langle h_{ij}\rangle$, where $h_{ij}$ is represented as a two-site MPO.
    (b) Approximate contraction for $\langle h_{ij}\rangle$ using corner and edge environment tensors $\{C,T\}$ to represent the semi-infinite surroundings. The environment bond dimension $\chi$ controls the accuracy of this approximation. The superscript $[i]$ labels the environment tensors associated with site tensor $A_i$.
    (c) Fixed-point equations defining the environment tensors. Gray triangles denote the isometries used for compression~\cite{Corboz2014,Fishman2018}. For brevity, we show only the equations for the upper-left corner and upper edge tensors; the remaining directions follow analogously.}
    \label{fig:CTMRG}
\end{figure}

Observables, in particular the single-particle Green's function $\langle \hat{c}^\dagger_i \hat{c}_j \rangle$ and density correlations $\langle \hat{n}_i \hat{n}_j \rangle$, are evaluated via the contractions of the corresponding infinite double-layer networks. Here, we use the directional CTMRG algorithm~\cite{Corboz2014, Fishman2018}, 
which iteratively finds a set of finite corner and edge environment tensors $\{C,T\}$ that approximates the semi-infinite surroundings of any finite rectangular patch of the double layer network supporting the observables [Fig.~\ref{fig:CTMRG}(a,b)]. Contracting these environments with the patch yields the desired expectation values, with an accuracy controlled by the \emph{environment} bond dimension $\chi$ of $\{C,T\}$; the exact results are recovered in the $\chi\to\infty$ limit.
Finally, to find the optimal iPEPS describing the ground state, we minimize the energy density $e_0$ via gradient-based L-BFGS optimization of the tensor elements in $\vec{A}$, combined with a strong Wolfe line search~\cite{NocedalWright2006}. Crucially, the gradient $\mathrm d e_0 /\mathrm d \vec{A}$ is computed via the fixed-point method~\cite{Liao2019,Ponsioen2022,Anna2025}, eliminating memory bottleneck in plain backpropagation due to $O(10^3)$ CTMRG steps required for convergence. 
For further technical details of the tensor network methods used in this work, we refer to Appendices~A and~B.

\begin{figure}[t]
    \centering
    \includegraphics[width=\columnwidth]{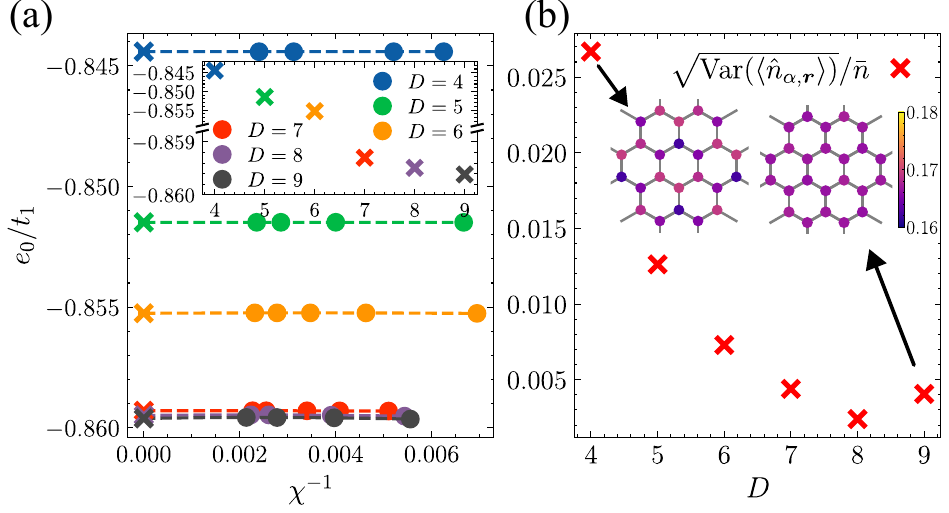}
    \caption{(a) Ground-state energy density $e_0(D,\chi)$ of the optimized iPEPS for bond dimensions from $D=4$ to $D=9$ and its extrapolations to $\chi\to\infty$. The extrapolated values are  plotted in the inset as a function of $D$.
    (b) Variance of the real-space charge distribution of the optimized states. For comparison, we also show the corresponding charge profiles $\{\langle \hat{n}_{\alpha, \boldsymbol{r}}\rangle\}$ for $D=4$ and $D=9$, plotted with a common color scale.}
    \label{fig:E_scaling}
\end{figure}

\emph{Results---}We optimize the $U(1)$-symmetric iPEPS up to $D=9$, choosing an identical $U(1)$ structure on all virtual spaces $\mathbb{V}_{t,l,b,r}=\oplus_q \mathbb{V}_q$, which we summarize in Table~\ref{tab:Dq-chi}. We report representative convergence and runtime characteristics in the Supplementary Material~\cite{si}.
First, we focus on the energetics of the optimized iPEPS, shown in Fig.~\ref{fig:E_scaling}(a),  together with the extrapolations of the ground state energy density to the $\chi \to \infty$ limit, which 
obey variational principle, serving as strict upper bounds on the true ground-state energy density in the thermodynamic limit.
As shown in the inset, the energy decreases substantially when increasing bond dimension from $D=4$  to $D=7$. Further increasing $D$ yields only a modest improvement. Moreover, by comparing with the ED results on finite torus clusters (up to 27 unit cells), we find that the iPEPS energy for $D\leq6$ lies above the first excitation gap, indicating strong competition among states within the variational manifolds of $D\leq6$, while $e_0$ of $D\geq 7$ are all well below the first excited gap. Moreover, in~\cite{si}, we compare the ground-state energies obtained from iPEPS and ED and find that the iPEPS energy extrapolated to $D\to\infty$ is consistent with the thermodynamic-limit extrapolation of the ED data. In addition to energy,
we examine the real-space charge inhomogeneity for increasing $D$. The inhomogeneity is quantified by the variance of the local density within the iPEPS unit cell [Fig.~\ref{fig:E_scaling}(b)],
\begin{equation}
    \text{Var} (\langle \hat{n}_{\alpha, \boldsymbol{r}}\rangle) = \frac{\sum_{\boldsymbol{r},\alpha=\text{A,B}}  (\langle  \hat{n}_{\alpha, \boldsymbol{r}}  \rangle - \bar{n})^2}{N},
\end{equation}
where the sum runs over honeycomb lattice sites in the iPEPS unit cell, $N$ is the number of sites in the iPEPS unit cell ($N=18$ in our case), and $\bar{n} = 1/6$ (per site) for $\nu=1/3$.
We find that states with $D<7$ exhibit significantly stronger charge inhomogeneity. Moreover, although clear chiral features already emerge in the entanglement spectrum for $D\geq4$ (qualitatively different from $D<4$), the exact counting structures for $4 \leq D<7$ (except for $D=4$) are not in quantitative agreement with the chiral $U(1)$ boson level counting (see~\cite{si}), unlike those for $D\geq7$. Taken together, these diagnostics imply a minimal bond dimension ($D_{\mathrm{min}}=7$ in our case) required to accurately capture the ground state and its chiral topological order, as also observed in bosonic cases~\cite{chen2018non,Sen2022}. For $D<7$, the variational manifold instead favors states with charge modulations. We therefore restrict the remainder of the paper to states with $D\geq7$.

\begin{figure}[t]
    \centering
    \includegraphics[width=\columnwidth]{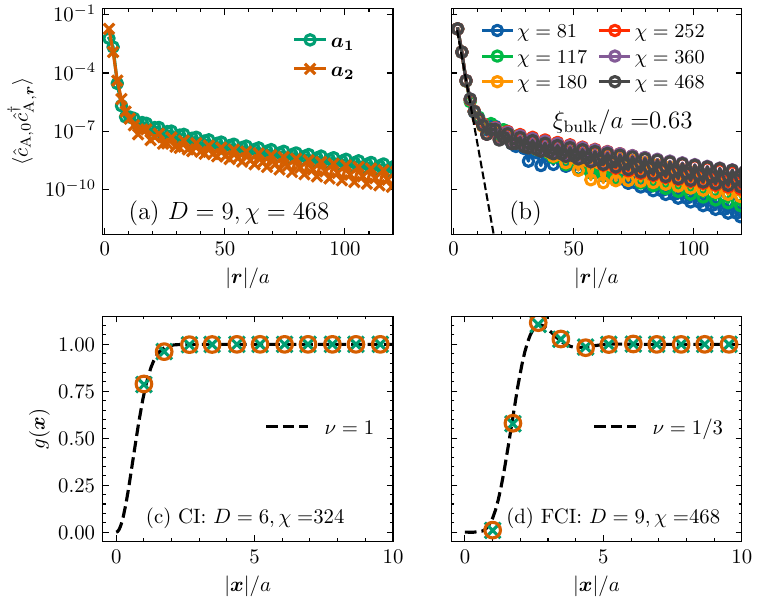}
    \caption{(a) Equal-time single-particle Green’s function $|\langle \hat{c}_{\mathrm{A}, 0}\hat{c}^{\dagger}_{\mathrm{A}, \boldsymbol{r}}\rangle|$ between A-A sublattice sites, evaluated along the two primitive lattice vector directions $\boldsymbol{a_1}$ and $\boldsymbol{a_2}$ for the $D=9$ state with $\chi=468$. (b) The Green's function along $\boldsymbol{a_2}$ of the $D=9$ state with increasing $\chi$.
    (c-d) Pair-correlation function $g(\boldsymbol{x})$ from iPEPS for the Chern insulator (CI) and the fractional Chern insulator (FCI), evaluated along the primitive lattice vectors $\boldsymbol{a_1}$ (cross) and $\boldsymbol{a_2}$ (circle). Here $|\boldsymbol{x}|$ is the Cartesian distance and $a$ is the honeycomb lattice constant. Dashed lines are pair-correlations in the continuum for the non-interacting $\nu=1$ IQH state and for the $\nu=1/3$ Laughlin state~\cite{Fulsebakke2023}, respectively, with $\ell_{B}/a=1.63$.
    }
    \label{fig:c_cp_corr}
\end{figure}

Next, we discuss the bulk properties captured by the correlation functions. In Fig.~\ref{fig:c_cp_corr}(a), we present the equal-time single-particle Green's function $\langle \hat{c}_{\text{A,0}}\hat{c}^\dagger_{\text{A}, \boldsymbol{r}}\rangle$, evaluated along the two primitive lattice vectors $\boldsymbol{{a}_1}$ and $\boldsymbol{a}_2$. We observe a similar behavior as found in CSLs~\cite{Poilblanc2017chiral, Juraj2022, Sen2022,Budaraju2024,chen2025simulating} and bosonic FQH states~\cite{Weerda2024FQHE}: a rapid exponential decay with a correlation length $\xi_{\mathrm{bulk}}/a=0.63$ ($a$ is the honeycomb lattice constant)---consistent with a gapped bulk---followed by an artifact of iPEPS representation, a long-range ``gossamer" tail (note the relatively tiny amplitude of the tail). The latter reflects the no-go theorem, prohibiting exact finite-$D$ iPEPS representation of states with chiral edge modes~\cite{Dubail2015}.
Increasing $\chi$ flattens the long-range tail but leaves the bulk part unchanged [see Fig.~\ref{fig:c_cp_corr}(b)]. In practice, the no-go theorem does not obstruct the identification of the key FCI signatures.

The pair-correlation function $g(\boldsymbol{x})$ is another important quantity in FQH physics \cite{Girvin1984,Girvin1986}. It reveals information about the probability of finding two fermions at a relative distance $\boldsymbol x$. The exponentially converging behavior of $g(\boldsymbol{x})$ to unity as $|\boldsymbol{x}|/\ell_B \to \infty$ is a signature of incompressibility ($\ell_B$ is the magnetic length). At short distances $|\boldsymbol{x}|/\ell_B \to 0$, $g(\boldsymbol{x})$ exhibits a correlation hole, reflecting the generalized exclusion principle of FQH states.
On the lattice, $g(\boldsymbol{x})$ is defined as
\begin{equation}
    g(\boldsymbol{x}) = \frac{\langle\psi| \hat{n}_0 \hat{n}_{\boldsymbol{x}} |\psi\rangle}{\langle \psi | \hat{n}_0|\psi\rangle\langle \psi| \hat{n}_{\boldsymbol{x}}|\psi\rangle},
\end{equation}
where $\hat{n}_{\boldsymbol{x}}$ is the local fermion number operator at site $\boldsymbol{x}$. On the honeycomb lattice, $\hat{n}_{\boldsymbol x}$ acts on either A or B sublattice depending on the position $\boldsymbol x$. In Fig.~\ref{fig:c_cp_corr}(d), we show the pair-correlation function of the FCI state with the reference site fixed to the B sublattice at the origin. Correlations along the two primitive lattice vectors $\boldsymbol {a_1}$ and $\boldsymbol{a_2}$ are shown and nearly coincide.
The function $g(\boldsymbol x)$ exhibits short-range oscillations and rapidly approaches unity.
For comparison, Fig.~\ref{fig:c_cp_corr}(c) shows $g(\boldsymbol x)$ for the corresponding CI obtained by simulating the model with $V_1=0$ at $\nu=1$. In this case $g(\boldsymbol x)$ approaches unity monotonically without oscillations.
Furthermore, we overlay continuum $g(\boldsymbol{x})$ for the Laughlin $\nu=1/3$ state (via the polynomial-expansion method \cite{Fulsebakke2023}) and for the non-interacting $\nu=1$ IQH state, $g(\boldsymbol x)=1-\exp(-|\boldsymbol x|^{2}/2\ell_B^{2})$ in the lowest Landau level. Remarkably, we find very good agreement with the lattice data for both cases using a single magnetic length $\ell_B/a \approx 1.63$.
This result underscores how even quantitative properties of these states transcend the particular microscopic realization. 

\setlength{\tabcolsep}{9pt} 
\begin{table}[t]
  \centering
  \caption{Optimized $U(1)$-symmetric iPEPS. Total bond dimension $D=\sum_q D_q$ and bond dimensions $D_q=\text{dim}(\mathbb{V}_q)$ of the individual charge-$q$ sectors, largest environment bond dimension $\chi_{\text{opt}}$ used during optimizations, and variational estimate of ground-state energy density $e_0(\chi \to \infty)$ obtained by extrapolations in $\chi$ [see Fig.~\ref{fig:E_scaling}(a)].}
  \begin{tabular}{c c c l}
    \hline\hline
    $D$ & $\{D_{q}\}$ & $\chi_{\text{opt}}$ & $e_0(\chi\to\infty)$\\
    \hline
    4 & $(D_{-1}, D_0, D_1) = (1, 2, 1)$ & 64 & -0.8444092(17) \\
    5 & $(D_{-1}, D_0, D_1) = (1, 2, 2)$ & 75 & -0.8514863(5) \\
    6 & $(D_{-1}, D_0, D_1) = (2, 2, 2)$ & 72 & -0.855254(5) \\
    7 & $(D_{-1}, D_0, D_1) = (2, 3, 2)$ & 98 & -0.8592989(18) \\
    8 & $(D_{-1}, D_0, D_1) = (2, 3, 3)$ & 112 & -0.859490(4) \\
    9 & $(D_{-1}, D_0, D_1) = (3, 3, 3)$ & 117 & -0.859614(12)  \\
    \hline\hline
  \end{tabular}
  \label{tab:Dq-chi}
\end{table}

Turning to truly universal properties, chiral topological order of FCIs can be characterized by their gapless edge theory, which, in our case, is a chiral $U(1)$ boson theory. 
Importantly, due to the Li–Haldane correspondence~\cite{Li2008,Qi2012Entanglement}, we do not need to impose open boundaries to create a physical edge,
instead, the low-lying edge spectrum is obtained from the entanglement spectrum (ES) of the ground state, which is readily accessible within iPEPS via the bulk–edge correspondence \cite{Cirac2011}. 
More specifically, we use the optimized iPEPS $|\psi\rangle$ to construct a PEPS wave function $|\Psi\rangle$ on an infinite horizontal cylinder of width $W$ (in unit cells) by tiling the cylinder with iPEPS unit cells. Under the bipartition of the cylinder into left and right halves, the entanglement Hamiltonian $H_{\text{ent}}$ is defined as the logarithm of the reduced density matrix for the left part $H_{\text{ent}} = -\ln (\rho_{\text{L}})$ with $\rho_{\text{L}}=\text{Tr}_{\text{R}}|\Psi\rangle\langle\Psi|$ (or analogously for the right half). As shown in Ref.~\cite{Cirac2011}, the eigenvalues of $\rho_{\text{L}}$ can be calculated using the following relation
\begin{equation}
    \text{spec}(\rho_{\text{L}}) = \text{spec}(\sqrt{\sigma_{\text{L}}^T} \sigma_{\text{R}} \sqrt{\sigma_{\text{L}}^T}) = \text{spec}(\sigma_{\text{L}}^T \sigma_{\text{R}}),
\end{equation}
where $\text{spec}(A)$ denotes the non-zero spectrum of the matrix $A$, and $\sigma_{\text{L}}$ and $\sigma_{\text{R}}$ are matrices formed from the left and right leading eigenvectors of the column-to-column transfer matrix of the double-layer norm $\langle \Psi | \Psi\rangle$ network. Since we use a $3\times 3$ iPEPS unit cell, the transfer matrix consists of three columns.
The matrices $\sigma_{\mathrm L}$, $\sigma_{\mathrm R}$, and their product $\sigma_{\mathrm L}^{T}\sigma_{\mathrm R}$ are approximated by the
matrix–product operators (MPOs) constructed from the CTMRG environment tensors, as illustrated in Fig.~\ref{fig:coarse_grain}(a,b). Note that swap gates are introduced to encode fermionic statistics, which can be subsequently absorbed into the MPO tensors. For a fermionic PEPS on a finite-width cylinder one may impose anti-periodic (APBC) or periodic (PBC) boundary conditions along the circumference, depending on the definition of the lattice translation symmetry (see \cite{si}). In Fig.~\ref{fig:coarse_grain}(b) we have implicitly assumed APBC.

\begin{figure}[t]
    \centering
    \includegraphics[width=\columnwidth]{Figs/ent_compress.pdf}
    \caption{(a) Left/right dominant eigenvectors $\sigma_{\mathrm L}$ and $\sigma_{\mathrm R}$ of the column-to-column double-layer transfer matrix. For clarity, we omit the arrow directions on tensor legs. The eigenvectors are approximated by MPOs assembled from the CTMRG environment edge tensors $T_{\mathrm L/\mathrm R}^{[i]}$; see Fig.~\ref{fig:CTMRG} for the definitions of the double-layer tensor and the environment tensors.
    (b) MPO representation of $\sigma_{\mathrm L}^{T}\sigma_{\mathrm R}$ on a cylinder of width $W=2$. The MPO for $\sigma_{\mathrm L}^{T}\sigma_{\mathrm R}$ has a three-tensor unit cell; small squares indicate fermionic swap gates.
    (c) MPO unit-cell compression for a 3-site unit cell: at each step, isometries $P_{\mathrm L}^{(i)}$ and $P_{\mathrm R}^{(i)}$ block two neighboring sites within the unit cell, reducing the unit-cell size by one while truncating the MPO ``physical" leg to dimension $\le D_{\mathrm c}$.}
    \label{fig:coarse_grain}
\end{figure}

Directly solving for the spectrum of the periodic MPO $\sigma_{\text{L}}^T \sigma_{\text{R}}$ quickly becomes infeasible as the cylinder width $W$ increases, especially for large MPO unit cells: The Hilbert-space dimension grows as $O(D^{W M})$, where $D$ is the ``physical" dimension of the MPO (equal to the iPEPS bond dimension), and $M$ is the MPO unit-cell size. We therefore introduce a pre-compression step before diagonalization. The idea is to coarse-grain the MPO unit cell to a single site, reducing the dimension from $D^{M}$ to  an effective dimension $D_{\text{c}}$. 
The procedure is summarized in Fig.~\ref{fig:coarse_grain}(c). At step-$i$, we apply isometries $P_{\text{L/R}}^{(i)}$ to two neighboring sites within the unit cell and obtain a single site with a dimension bounded by $D_{\text{c}}$. The iteration is repeated until the unit cell is coarse grained into a single site.
The construction of $P^{(i)}_{\text{L/R}}$ follows Ref.~\cite{Xie2012}, which utilizes the higher-order singular value decomposition (HOSVD). For simplicity, we always construct the isometries based on the HOSVD of $\sigma_{\text{L}}$ and then apply the isometries to both $\sigma_L$ and $\sigma_R$.

\begin{figure}[t]
    \centering
    \includegraphics[width=\columnwidth]{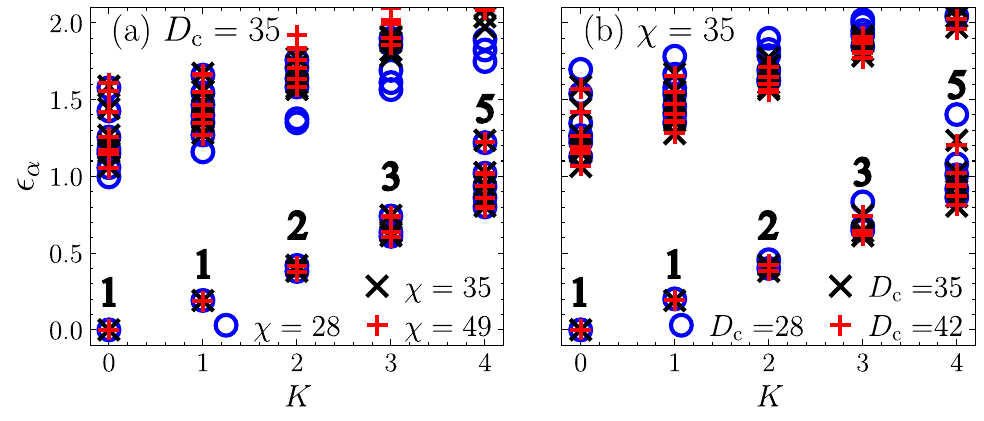}
    \caption{Momentum-resolved ES of the $D=7$ state on a cylinder of circumference $W=5$ (15 sites) under APBC. The spectrum is calculated for the $U(1)$ charge sector $n=0$. (a) The cutoff bond dimension $D_{\mathrm{c}}=35$ is fixed, and the environment bond dimension $\chi$ is varied. (b) The environment bond dimension $\chi=35$ is fixed, and the cutoff bond dimension $D_{\mathrm{c}}$ is varied. In both cases, the low-level counting across momentum sectors follows $(1, 1, 2, 3, 5)$, consistent with the chiral $U(1)$ boson edge of the $\nu=1/3$ Laughlin state.}
    \label{fig:edge_spec}
\end{figure}

With the pre-compression step, we are able to push the cylinder width to $W=5$ (15 sites) for the $D=7$ iPEPS. Figure~\ref{fig:edge_spec} shows the resulting momentum-resolved ES in the $U(1)$ charge sector $n=0$ under APBC and compares the spectra obtained at several environment bond dimensions $\chi$ and MPO cutoff dimensions $D_{\text{c}}$. Remarkably, despite coarse-graining an MPO unit cell of dimension $D^M=7^3=343$ to a single site of dimension $D_{\text{c}}\approx28\text{–}42$, the low-energy levels are robust across these parameters and are already well converged even for a modest $\chi$ and $D_{\text{c}}$ (e.g., $\chi=28$ and $D_{\text{c}}=35$). Most importantly, we observe the characteristic level counting $(1,1,2,3,5)$ for $W=5$, providing a sharp fingerprint of the FCI state~\cite{Li2008,Qi2012Entanglement}. 
The same chiral $U(1)$ boson level counting is obtained under PBC, and it persists for the $D=8$ and $D=9$ states (see \cite{si}).

\emph{Discussion---}By combining state-of-the-art techniques, we have demonstrated that iPEPS can simulate chiral topological order in \emph{fermionic} systems by variationally realizing an FCI at filling $\nu=1/3$. In the bulk, the equal-time single-particle Green’s function shows a rapid exponential decay followed by a long-range ``gossamer” tail---consistent with the finite-$D$ iPEPS manifestation of the no-go constraint—and the pair-correlation function $g(\boldsymbol x)$ matches qualitative features of the $\nu=1/3$ Laughlin state, including the correlation hole and the rapid convergence to unity. The edge ES exhibits the characteristic $(1,1,2,3,5)$ counting. It is already well-converged at modest environment bond dimension $\chi$ and compression cutoff $D_{\text{c}}$, and remains robust to the choice of APBC/PBC along the cylinder. 

Our methodology readily extends to other filling fractions and lattice geometries, which would be useful in studying states in the hierarchy of composite fermion states and non-Abelian FCIs. The remaining computational bottleneck in simulating chiral topological order is the slow convergence of CTMRG in the forward pass (the backward pass in the fixed-point approach typically converges much faster). A fast algorithm for determining the environment tensors is highly desired.
In this work we fix the fractional filling by using a $U(1)$-symmetric ansatz and a multi-site iPEPS unit cell. Since FCI states do not break translation symmetry, a natural question is whether a single-site iPEPS could approximate a fermionic FCI by relaxing $U(1)$ to $\mathbb Z_2$ and tuning the density via a chemical potential. Single-site iPEPS are easier to optimize and can be further endowed with lattice symmetries. 
In addition to the ground state, another interesting direction is to apply recently developed iPEPS excitation ansatz \cite{Vanderstraeten2019,Ponsioen2020,Boris2022,Tu2024} to compute neutral and charged excitations on top of the FCI state, providing dispersion relations and gaps directly in the thermodynamic limit.

Our work adds iPEPS to the few numerical methods that can be used to simulate chiral fermionic topological order, in a time where experimental realizations of such exotic phases call for quantitatively reliable predictions.

\emph{Acknowledgments---}We thank Marek Rams, Anna Fracuz, Paul Brehmer, Johannes Motruk, and Sen Niu for valuable discussions. Tensor network simulations were performed with open-source library \textsc{YASTN} \cite{Marek2025yastn}.

The authors acknowledge support from the Swiss National Science Foundation through a Consolidator Grant (iTQC, TMCG-2213805) and a Quantum grant (20QU-1225225). 
This project further received financial support from Google. We thank NVIDIA and the MesoNET project (allocation M25112, Turpan cluster) for providing access to GPU computing resources that supported this work.

\emph{Data availability---}The data that support the findings of this work are openly available~\cite{chen2026fci}.

\bibliography{ref}

\onecolumngrid
\begin{center}
  \section*{End Matter}
\end{center}
\twocolumngrid
\emph{Appendix~A: iPEPS ansatz details---}We impose $U(1)$-symmetry on iPEPS tensors to reduce computational cost and to fix the fermion filling. 
The vector spaces associated to $A^{[\boldsymbol{r}]}$ are then decomposed into direct sums of representations of the $U(1)$ group $\mathbb V =\bigoplus_{q\in\mathbb Z} \mathbb V_q$, where $\mathbb V_q$ satisfies
\begin{equation}\label{eq:U1_sector}
\quad \forall \boldsymbol{w} \in \mathbb V_q, \,g=\exp(i\theta \hat Q) \in U(1): g\cdot \boldsymbol w = e^{i\theta q}\boldsymbol w.
\end{equation}
As an example, the local physical Hilbert space is decomposed as $\mathbb V_{\mathrm{p}}=\mathbb V_0\oplus \mathbb V_1 \oplus \mathbb V_2$
for spinless fermions, corresponding to occupations $n=0,1,2$ within the Bravais lattice unit cell. 
Similarly, each virtual leg is split into sectors
\begin{equation}
\mathbb V_{\mathrm{virt}}=\bigoplus_{q \in \mathbb Z} \mathbb V_q,\qquad \dim \mathbb V_q = D_q .
\end{equation}
We call $A^{[\boldsymbol r]}$ $U(1)$-symmetric if it transforms as a singlet under $g=\exp(i\theta \hat Q)$:
\begin{equation}\label{eq:U1-transformation}
\begin{aligned}
(g \cdot A^{[\boldsymbol r]})_{i_t i_l i_b i_r i_p}
&:= e^{i\theta\,(q_{i_t}+q_{i_l}-q_{i_b}-q_{i_r}+q_{i_p})}
A^{[\boldsymbol r]}_{i_t i_l i_b i_r i_p} \\
&\stackrel{!}{=} e^{i\theta n}\,A^{[\boldsymbol r]}_{i_t i_l i_b i_r i_p},
\end{aligned}
\end{equation}
where $q_\alpha$ is the $U(1)$ charge carried by index $\alpha$, and $n$ is the total charge of $A^{[\boldsymbol{r}]}$. Eq.~\eqref{eq:U1-transformation} implies that $A^{[\boldsymbol r]}$ is block-sparse, and the nonzero blocks occur only when
\begin{equation}
    q_{i_t}+q_{i_l}+q_{i_p}-q_{i_b}-q_{i_r} = n.
\end{equation}
In practice we fix $n=0$ for all iPEPS tensors $A^{[\boldsymbol{r}]}$. The block-sparse structure enables optimizations at larger bond dimensions than in the dense case. 

The $U(1)$ symmetry is also helpful in fixing the filling of fermions. As noted in Refs.~\cite{Bauer2011symmetry,bruognolo2021beginner}, an iPEPS built from $U(1)$-symmetric tensors with $n=0$ can only represent states in the global $q=0$ sector—i.e., zero fermion density. To target finite fillings, we attach an auxiliary leg to $A^{[\boldsymbol r]}$, which carries a single charge sector $\mathbb V_{-q_0}$ with $\dim \mathbb V_{-q_0}=1$. Fusing this leg with the physical leg shifts the charge label by $-q_0$
\begin{equation}
    \mathbb V_q \otimes \mathbb V_{-q_0} = \mathbb{V}_{\tilde q = q-q_0}.
\end{equation}
If we denote $Q_0 = \sum_{\boldsymbol r} q_0 \in \mathbb{Z}$ the sum of attached charges to tensors within the iPEPS unit cell, then the representable states in the new global $\tilde q=0$ sector correspond to physical states at filling $\nu=Q_0/(L_x L_y)$, with $L_x L_y$ the unit-cell size of the iPEPS. Therefore, to impose fractional filling, multi-site unit cell is needed. In this work we take a $3\times 3$ iPEPS unit cell with three independent tensors, as shown in Fig.~\ref{fig:model}(b). We attach an auxiliary charge $q_0=1$ only to tensor $A_1$ so the filling is fixed to $\nu=1/3$.

We adopt the swap-gate approach to encode fermionic statistics \cite{Corboz2009,Corboz2010}. In this framework, each leg is decomposed into even (bosonic)  and odd (fermionic) parity sectors $\mathbb V=\mathbb V_+\oplus\mathbb V_-$, assuming the system has the $\mathbb Z_2$ parity symmetry. 
Because $\mathbb {Z}_2$ is a subgroup of $U(1)$, the $U(1)$-charge sector $\mathbb V_q$ of Eq.~\eqref{eq:U1_sector} can be labeled by parity. Specifically, we have 
\begin{equation}
    \mathbb V_+= \bigoplus_{q \text{ even}} \mathbb V_q, \ 
    \mathbb V_-= \bigoplus_{q \text{ odd}} \mathbb V_q.
\end{equation}
A rank-4 tensor called the swap-gate is introduced, which captures the swapping process between bosonic/fermionic modes:
\begin{equation}\label{eq:swap-gate}
    \includegraphics[width=.6\columnwidth]{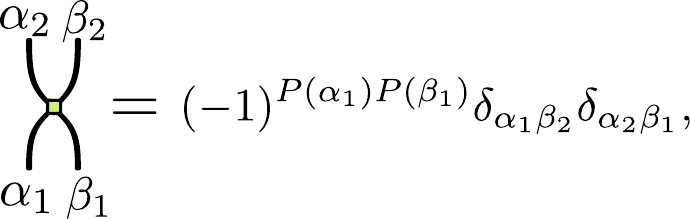}
\end{equation}
where $P(\alpha)$ takes value on 0 or 1, representing the parity of index $\alpha$.  Eq.~\eqref{eq:swap-gate} yields a minus sign only when both incoming legs are fermionic $(P(\alpha_1)=P(\beta_1)=1)$. 
Every line crossing in a tensor diagram is decorated with a swap gate. This enforces the correct fermionic signs while leaving the asymptotic contraction complexity unchanged.

\emph{Appendix~B: CTMRG and fixed-point AD--}As mentioned in the main text, CTMRG produces a set of environment tensors $\vec{E}=\{C,T\}$ used to evaluate local observables. These tensors approximate the semi-infinite surroundings of the rectangular patch of the double-layer network supporting a given observable, as illustrated in Fig.~\ref{fig:CTMRG}(a,b) for the $3\times 3$ unit cell employed in the main text. From this perspective, it is natural to formulate a set of fixed-point equations that characterize $\vec{E}$, as shown in Fig.~\ref{fig:CTMRG}(c). Importantly, due to the internal gauge freedom on the environment virtual legs, these relations are satisfied only up to invertible gauge transformations~\cite{Liao2019,Boris2022,naumann2024introduction,Anna2025}, indicated by the symbol ``$\sim$'' in Fig.~\ref{fig:CTMRG}(c).

In practice, we obtain the environment tensors by iterating the fixed-point equations and monitor CTMRG convergence via the singular-value spectra of the corner tensors $C$, which are gauge invariant. Since we do not impose gauge fixing, the resulting converged environment is a fixed point only up to gauge transformations. This poses no difficulty for conventional AD, as the gradient $\mathrm d\vec{E}/\mathrm d\vec{A}$ is obtained by backpropagating through the full CTMRG iteration history. However, the gauge freedom becomes an obstruction for the fixed-point AD method. Fixed-point AD relies on the assumption that $\vec{E}$ satisfies \emph{strict} fixed-point equations, i.e., that the relations in Fig.~\ref{fig:CTMRG}(c) hold with ``$\sim$'' replaced by ``$=$''. Denoting this strict fixed-point condition by $\vec{E}=f(\vec{E},\vec{A})$ and writing $\vec{E}^*(\vec{A})$ for its solution, the gradient $\mathrm d\vec{E}/\mathrm d\vec{A}$ can be expressed as
\begin{equation}\label{eq:fixed_pt_AD}
\begin{aligned}
\left.\frac{\mathrm d \vec{E}}{\mathrm d \vec{A}}\right|_{\vec{E}=\vec{E}^*}
&=\left(1-\left.\frac{\partial f}{\partial \vec{E}}\right|_{\vec{E}=\vec{E}^*}\right)^{-1}
\left.\frac{\partial f}{\partial \vec{A}}\right|_{\vec{E}=\vec{E}^*} \\
&=\sum_{n=0}^{\infty}
\left(\left.\frac{\partial f}{\partial \vec{E}}\right|_{\vec{E}=\vec{E}^*}\right)^{n}
\left.\frac{\partial f}{\partial \vec{A}}\right|_{\vec{E}=\vec{E}^*}.
\end{aligned}
\end{equation}
A key advantage of computing the gradient via Eq.~\eqref{eq:fixed_pt_AD} is that it requires only the local Jacobians evaluated at the fixed point $\vec{E}^*$, thereby avoiding the need to store the full CTMRG iteration history as in conventional AD. 

\begin{figure}[tb]
    \centering
    \includegraphics[width=\columnwidth]{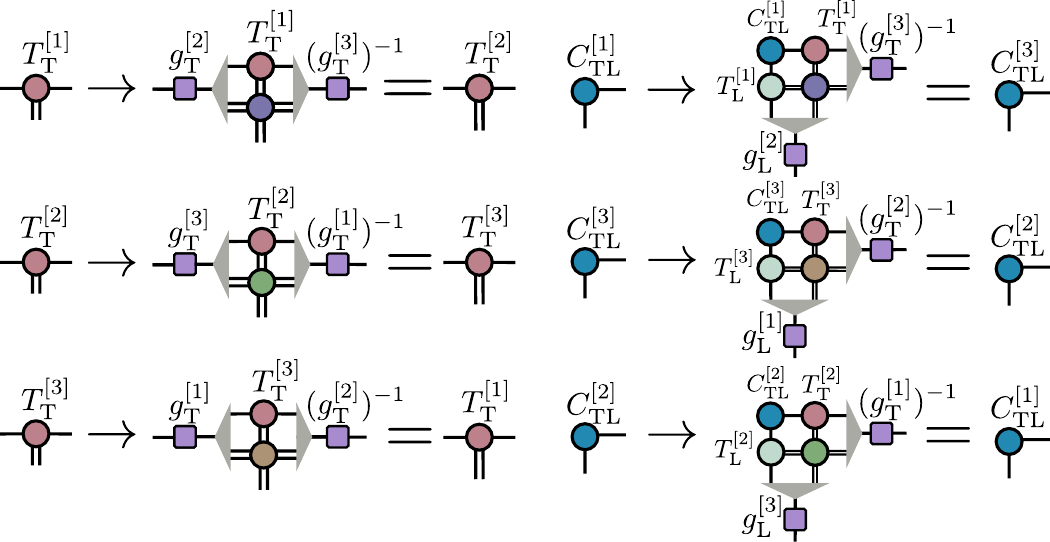}
    \caption{Action of the gauge-fixed map $G\circ f$. The gauge transformation $G$ is parametrized by a set of  invertible matrices $\mathcal{G}=\{g^{[i]}_{\mathrm{T/L/B/R}}\}$.
    For brevity, we show only the map for the upper-left corner and upper edge tensors; the remaining directions follow analogously.}
    \label{fig:gauge_fixing}
\end{figure}

Therefore, to implement fixed-point AD one must fix the gauge by finding an invertible transformation $G$ such that the CTMRG output $\vec{E}_0$---which is not a strict fixed point of $f$---becomes an elementwise-converged fixed point of the gauge-fixed map $G\circ f$:
\begin{equation}
   \vec{E}_0 = G\circ f(\vec{E}_0, \vec{A}).
\end{equation}
The action of the gauge-fixed map $G\circ f$ is illustrated in Fig.~\ref{fig:gauge_fixing}, where we introduce a set of invertible matrices $\mathcal{G}=\{g^{[i]}_{\mathrm{T/L/B/R}}\}$. The number of independent matrices $g^{[i]}_{\mathrm{T/L/B/R}}$ matches the number of edge tensors. We follow the scheme of Ref.~\cite{Anna2025} to construct $\mathcal{G}$. In our implementation, gauge fixing is invoked only after CTMRG has converged the environment (up to gauge transformations). Moreover, $\mathcal{G}$ is treated as a post-processing gauge choice and detached from the computation graph: since it acts only by invertible gauge transformations on the environment legs, detaching $\mathcal{G}$ does not affect gradients of gauge-invariant observables. With this gauge-fixing procedure in place, fixed-point AD can then be applied.

\onecolumngrid          
\clearpage
\appendix
\section*{Supplemental Material}

\input{SM}

\end{document}

%% file: SM.tex
\title{Supplementary Material - Simulating fermionic fractional Chern insulators \\
with infinite projected entangled-pair states}

\maketitle

\setcounter{figure}{0}
\setcounter{equation}{0}
\setcounter{table}{0}
\renewcommand\thefigure{A\arabic{figure}}
\renewcommand\theequation{A\arabic{equation}}
\renewcommand\thetable{A\arabic{table}}
\section{iPEPS optimization history}

\begin{figure}[b]
    \centering
    \includegraphics[width=0.9\linewidth]{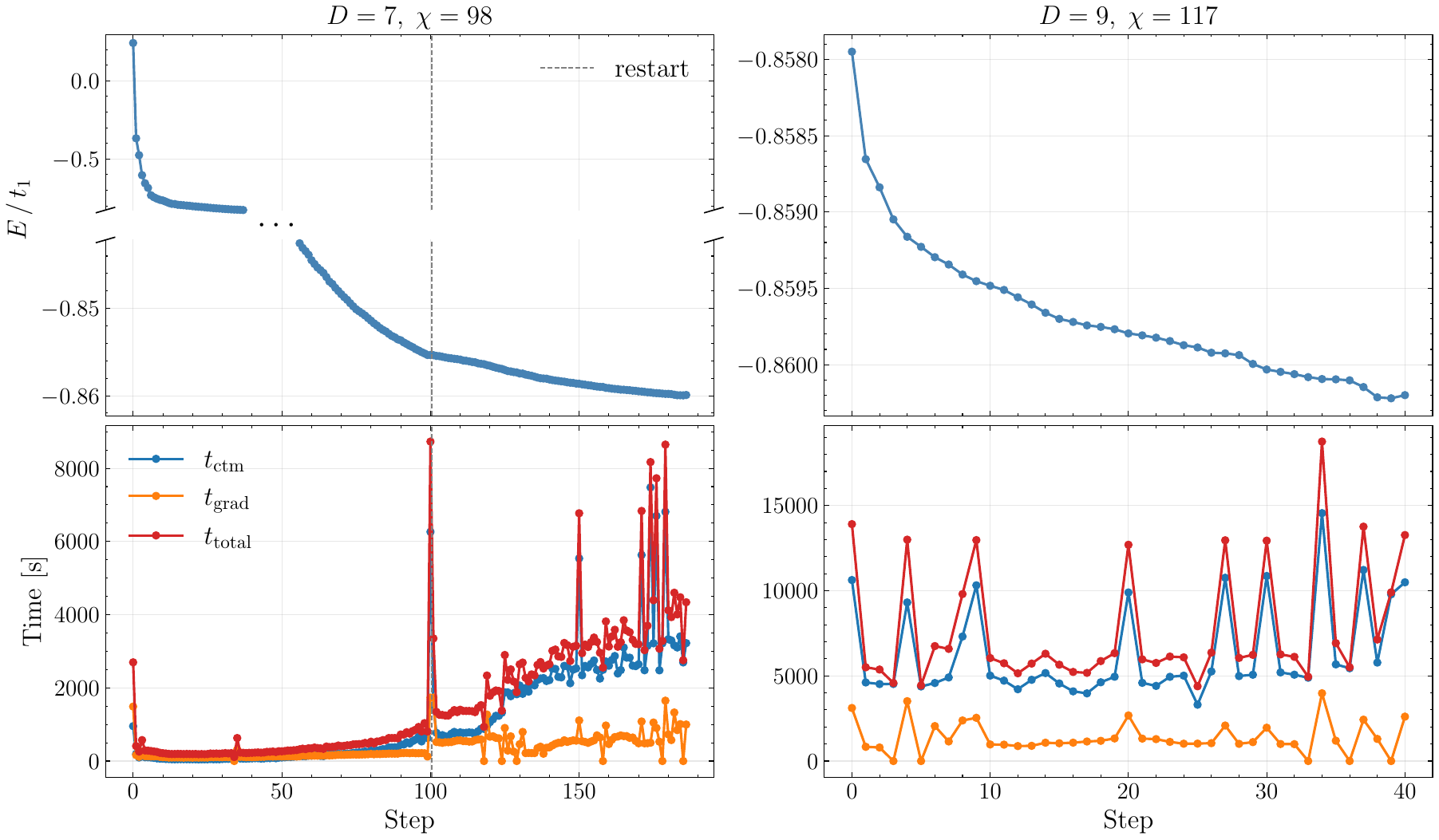}
    \caption{Optimization history for $(D,\chi)=(7,98)$ (left) and $(9,117)$ (right). The upper and lower panels show the energy and the runtime as functions of the optimization step, respectively. For $(D,\chi)=(7,98)$, we restart the optimization midway, indicated by the dashed vertical line. $t_{\mathrm{ctm}}$ and $t_{\mathrm{grad}}$ denote the time spent on CTMRG and on gradient evaluation in each optimization step.}
    \label{fig:optim_history}
\end{figure}

In this section, we report representative convergence and runtime characteristics in iPEPS optimizations.
The optimization histories for $(D,\chi)=(7,98)$ and $(9,117)$ are shown in Fig.~\ref{fig:optim_history}. At each optimization step, we perform a line search to determine the step size for the gradient update. The number of line-search trials varies from step to step because it depends on the local landscape of the loss function. This variability is also reflected in the runtime: some steps require multiple trials to identify an acceptable step size.

At step 100 in Fig.~\ref{fig:optim_history}(a), the line search returns an (effectively) vanishing step size after reaching the maximum number of trials. We attribute this to an inaccurate L-BFGS inverse-Hessian approximation, which can occasionally produce a poor search direction. In this case, we discard the stored L-BFGS history and restart the optimization, after which the energy continues to decrease. We also note that the runtime gradually increases as the state approaches the FCI regime. This is consistent with a growing correlation length, which increases the number of CTMRG iterations required to reach convergence.

Optimizing higher-$D$ states (e.g., $D=9$) from random initialization is computationally expensive. Instead, we initialize the $D=9$ run by embedding the optimized $D=7$ iPEPS tensor into bond dimension $D=9$ (padding additional matrix entries with zeros) and adding a small amount of noise to the tensor elements. The resulting optimization curve is shown in the right column of Fig.~\ref{fig:optim_history}. All simulations were performed on four NVIDIA A100 GPUs. The total wall-clock cost is $4\times 83.7\,\mathrm{h}=334.8\,\mathrm{GPU\cdot h}$ for $(D,\chi)=(7,98)$ and $4\times 88.4\,\mathrm{h}=353.6\,\mathrm{GPU\cdot h}$ for $(D,\chi)=(9,117)$.

\section{Comparison with the results of exact-diagnoalization}
\begin{figure}[t]
    \centering
    \includegraphics[width=0.6\linewidth]{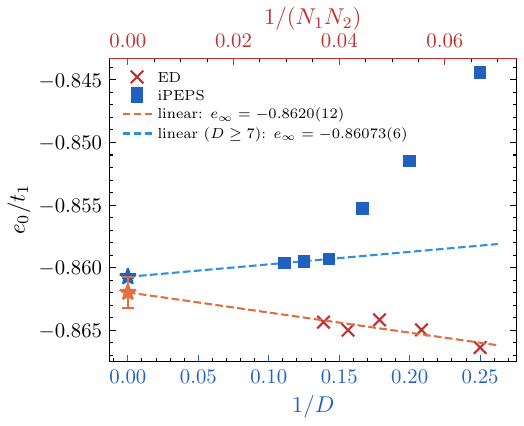}
    \caption{Comparison between iPEPS and exact-diagonalization ground-state energies.}
    \label{fig:ED_cmp}
\end{figure}

We compare the iPEPS energies extrapolated to $\chi \to \infty$ at each bond dimension $D$ with the ground-state energies obtained from exact diagonalization (ED) on finite tori; the results are shown in Fig.~\ref{fig:ED_cmp}. The largest ED cluster contains $N_1N_2=27$ honeycomb unit cells and $N_e=9$ spinless fermions. On the torus, the ground state exhibits an approximate threefold topological degeneracy; in Fig.~\ref{fig:ED_cmp}, we therefore plot the average energy of these three nearly degenerate states. To estimate the thermodynamic-limit ground-state energy, we perform a linear extrapolation of the iPEPS energies for $D\geq 7$ in $D^{-1}$, and a linear extrapolation of the ED energies in $(N_1N_2)^{-1}$. The resulting extrapolated energies from the two approaches are consistent with each other.

\section{APBC vs PBC on cylinders}
To resolve the entanglement spectrum in discrete  momenta, we require a ground state on a finite-width cylinder. Here we approximate it by embedding the optimized iPEPS tensors directly on a cylinder of circumference $W$.
For fermionic PEPS we can impose either antiperiodic (APBC) or periodic (PBC) boundary conditions along the circumference. The choice is encoded in the definition of the lattice translation operator $T_y$ around the cylinlder. Correspondingly, the resulting ground state should be invariant under $T_y$. In the following, we will specify the action of the translation operator $T_y$ on the double-layer transfer matrix\text{---}relevant for the entanglement spectrum\text{---}for both boundary conditions, and identify the corresponding $T_y$-invariant transfer matrices. 

To reduce clutter in the tensor-network diagrams we block the double-layer tensors within each iPEPS unit cell and represent the result as a single effective tensor.
For APBC, the lattice translation $T_{y}$ acting on the translation-invariant transfer matrix at width $W=3$ is defined as
\begin{equation}\label{eq:APBC_Ty}
    \includegraphics[width=.4\columnwidth]{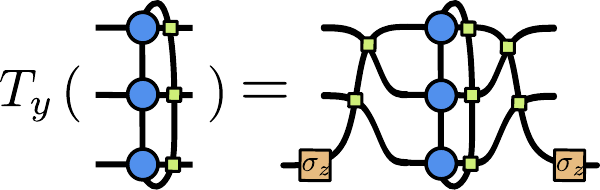}\quad,
\end{equation}
where the small squares denote the swap gate and $\sigma_z$ is the parity operator, defined as
\begin{equation}
   (\sigma_z)_{\alpha \beta} = 
   \begin{cases}
    -\delta_{\alpha \beta}, \text{ parity}(\alpha)=\text{odd}\\
    \delta_{\alpha\beta}, \text{ parity}(\beta) = \text{even}.
    \end{cases}
\end{equation}

The extra $\sigma_z$ is precisely what enforces APBC: when the degrees of freedom at site-$1$ are translated across the cut to site-$W$, odd-parity states acquire a minus sign. By using the jump-move \cite{Philippe2010}, it can be shown that the translated transfer matrix is equal to the original one:
\begin{equation}
    \includegraphics[width=0.5\columnwidth]{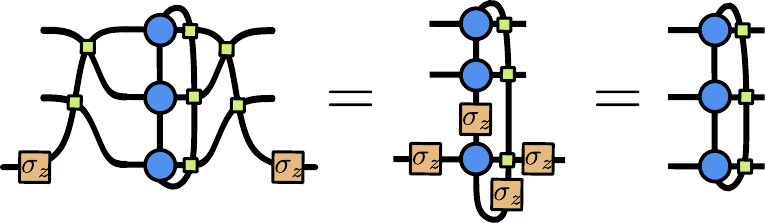}\quad,
\end{equation}
where in the second equality we used the fact that the double-layer tensor is parity invariant. Therefore, the transfer matrix in Eq.~\eqref{eq:APBC_Ty} is invariant under $T_y$ for APBC.

\begin{figure}[ht]
    \centering
    \includegraphics[width=.7\columnwidth]{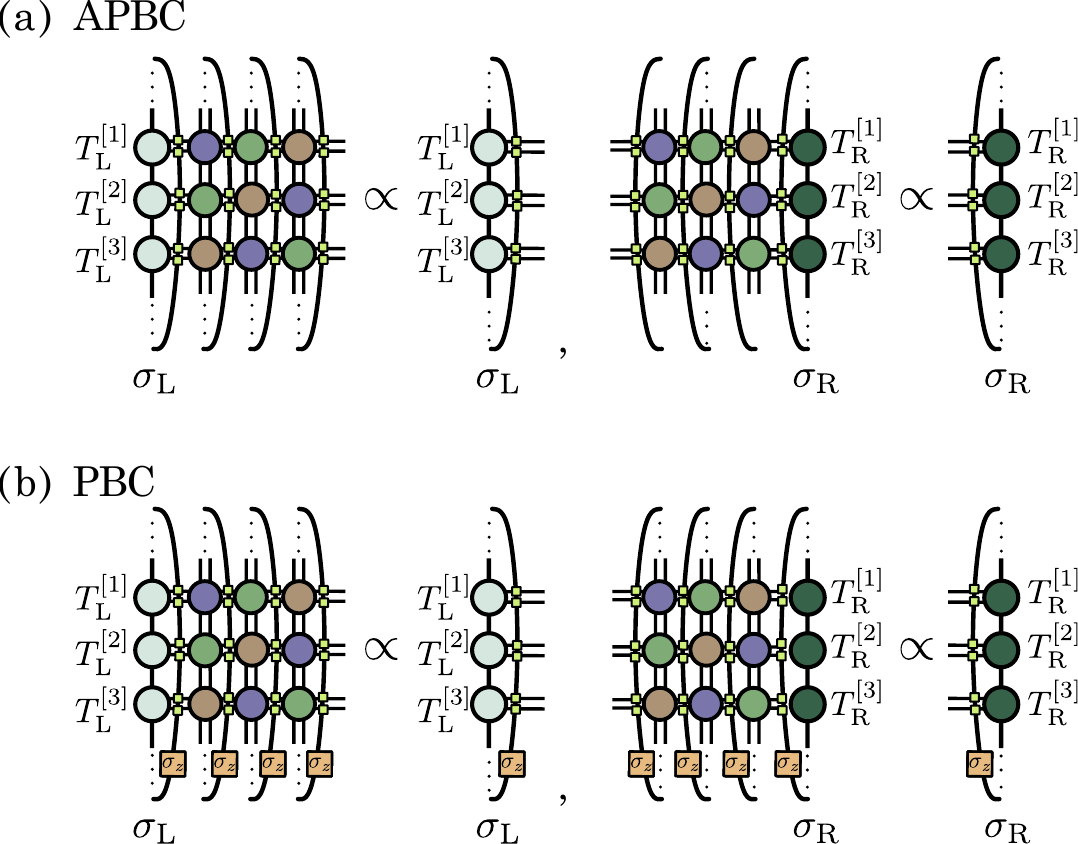}
    \caption{MPO approximations of the left and right leading eigenvectors, $\sigma_{\mathrm L}$ and $\sigma_{\mathrm R}$, of the double-layer transfer matrix (with a $3\times3$ unit cell) on a cylinder, for both APBC and PBC.}
    \label{fig:sigma_BCs}
\end{figure}

For PBC, the lattice translation $T_{y}$ acting on the translation-invariant transfer matrix at width $W=3$ is defined as 
\begin{equation}\label{eq:PBC_Ty}
    \includegraphics[width=0.4\columnwidth]{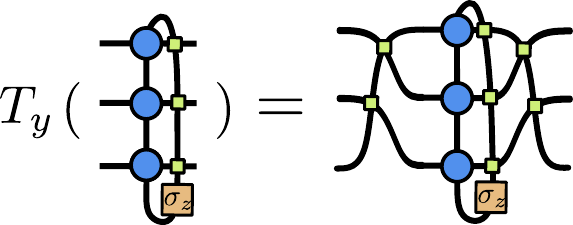}\quad.
\end{equation}
Note that we inserted a $\sigma_z$ operator when constructing the transfer matrix. Similarly, one can show the following identity
\begin{equation}
    \includegraphics[width=0.4\columnwidth]{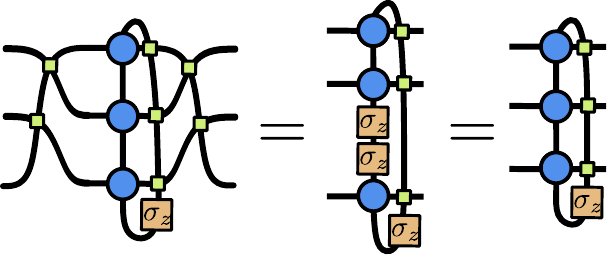}\quad .
\end{equation}
Therefore, the transfer matrix in Eq.~\eqref{eq:PBC_Ty} is invariant under $T_y$ for PBC.

\begin{figure}[htb]
    \centering
    \includegraphics[width=.7\textwidth]{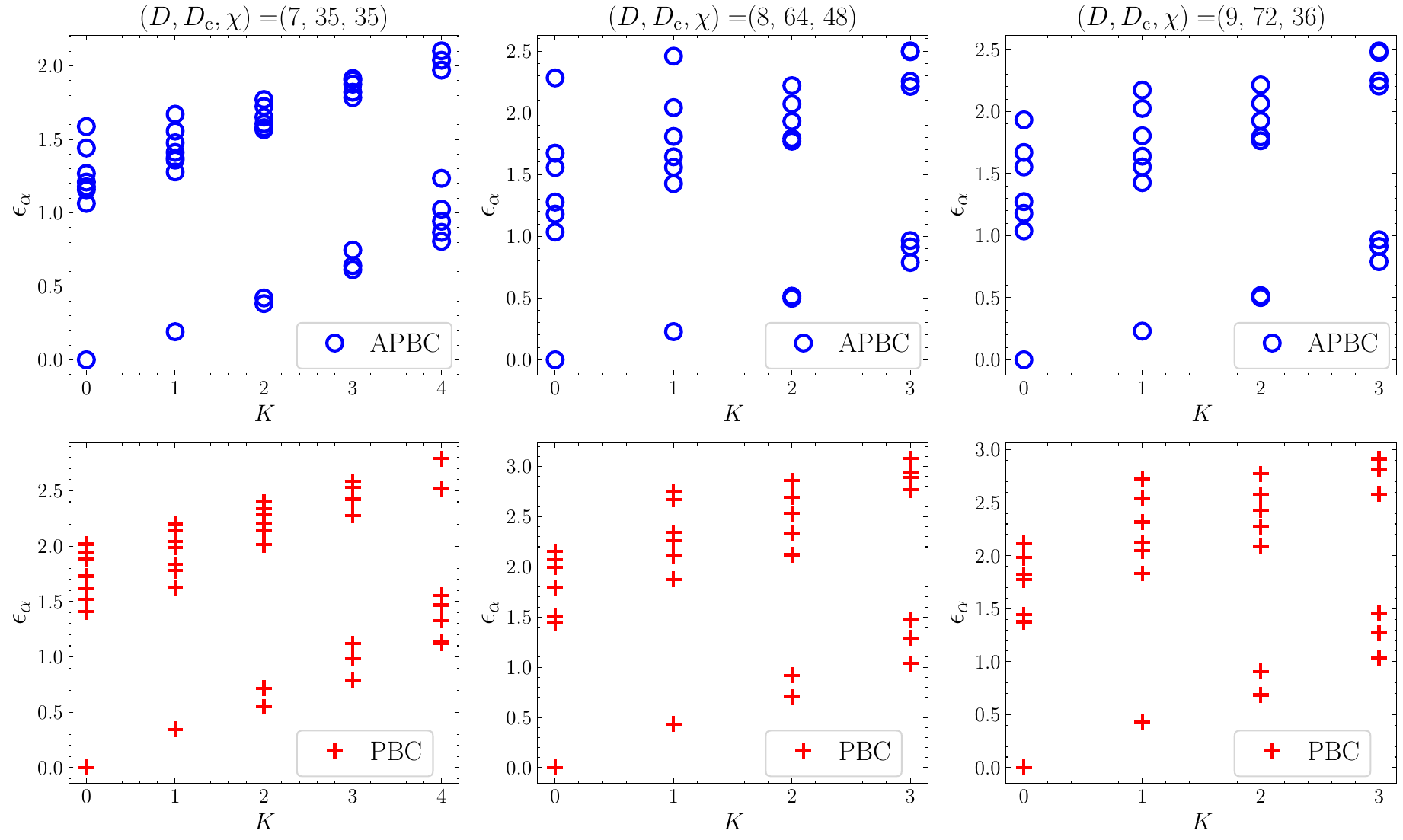}
    \caption{Momentum-resolved ES of the $D=7,8,9$ FCI states under APBC (top row) and PBC (bottom row). For each bond dimension $D$, the corresponding cutoff dimension $D_{\mathrm{c}}$ and environment bond dimension $\chi$ are indicated at the top of each column.}
    \label{fig:edge_spec_all}
\end{figure}

Using the environment edge tensors obtained from CTMRG, we construct the left and right leading eigenvectors, $\sigma_{\mathrm{L}}$ and $\sigma_{\mathrm{R}}$, of the transfer matrix under both APBC and PBC, as illustrated in Fig.~\ref{fig:sigma_BCs} for a $3\times 3$ iPEPS unit cell. We apply the compression technique introduced in the main text for both boundary conditions and compute the ES for $D=7$--$9$. As shown in Fig.~\ref{fig:edge_spec_all}, the chiral $U(1)$ boson level counting is consistently observed in all cases.

\begin{figure}[ht]
    \centering
    \includegraphics[width=0.9\linewidth]{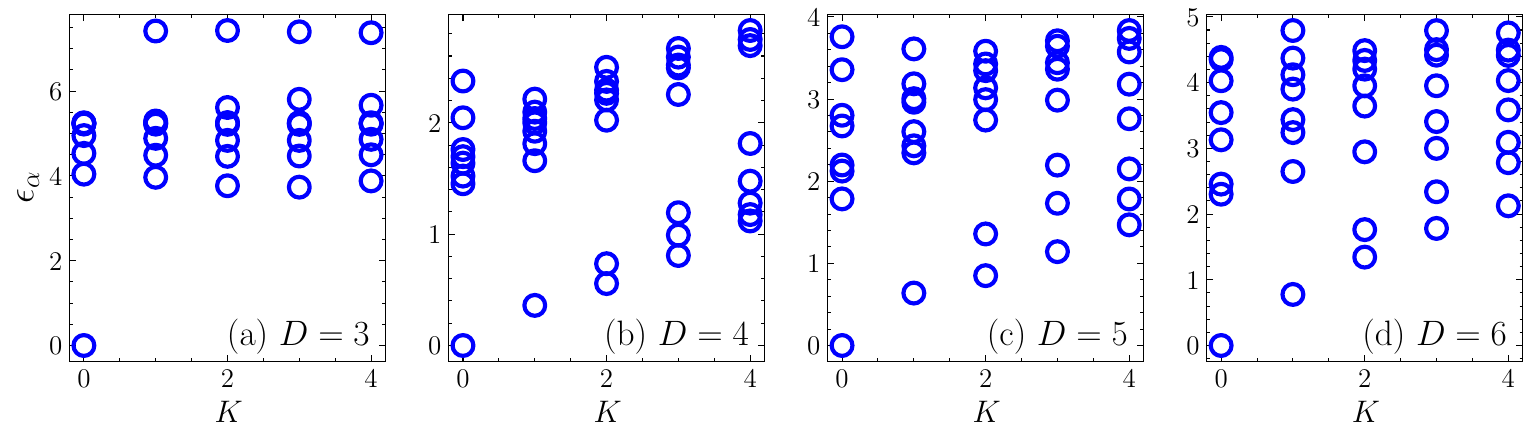}
    \caption{Momentum-resolved ES of the $D=3\text{--}6$  states under APBC.}
    \label{fig:es_3_to_6}
\end{figure}

For comparison, we also present in Fig.~\ref{fig:es_3_to_6} the ES for $D=3$--$6$ under APBC, computed using the same procedure. The ES at $D=3$ is qualitatively distinct from that at $D\ge 4$. Starting from $D=4$, the ES develops clear chiral features. For $D=5$ and $D=6$, the spectral levels  the spectral levels blend into the continuum at large momentum sectors $K$, obscuring the expected counting structure. While the $D=4$ state exhibits the correct counting, it shows substantial charge fluctuations (see the main text). We therefore identify $D=7$ as the minimal bond dimension required for a faithful representation of the FCI state.

